# Tailored flux pinning in superconductor/ferromagnet multilayers with engineered magnetic domain morphology – from stripes to skyrmions

X. Palermo[1], N. Reyren[1], S. Mesoraca[1], A. V. Samokhvalov[2], S. Collin[1], F. Godel[1], A. Sander[1], K. Bouzehouane[1], J. Santamaría[1,3], V. Cros[1], A.I. Buzdin[1,4] and J.E. Villegas[1*]

*[1]Unité Mixte de Physique, CNRS, Thales, Université Paris-Saclay, 91767, Palaiseau, France*
*[2] Institute for Physics of Microstructures, Russian Academy of Sciences, 603950 Nizhny Novgorod, GSP-105, Russia*
*[3] Grupo de Física de Materiales Complejos, Dpt. Física de Materiales, Universidad Complutense de Madrid, 28040 Madrid, Spain*
*[4] Université de Bordeaux, LOMA UMR CNRS 5798, F-33405 Talence, France*

Superconductor/Ferromagnet (S/F) hybrid systems show interesting magneto-transport behaviors that result from the transfer of properties between both constituents. For instance, magnetic memory can be transferred from the F into the S through the pinning of superconducting vortices by the ferromagnetic textures. The ability to tailor this type of induced behavior is important to broaden its range of application. Here we show that engineering the F magnetization reversal allows tuning the strength of the vortex pinning (and memory) effects, as well as the field range in which they appear. This is done by using magnetic multilayers in which Co thin films are combined with different heavy metals (Ru, Ir, Pt). By choosing the materials, thicknesses, and stacking order of the layers, we can design the characteristic domain size and morphology, from out-of-plane magnetized stripe domains to much smaller magnetic skyrmions. These changes strongly affect the magneto-transport properties. The underlying mechanisms are identified by comparing the experimental results to a magnetic pinning model.

* javier.villegas@cnrs-thales.fr

## I. INTRODUCTION

The interplay between superconductivity and magnetism in artificial hybrid structures [1] leads to a rich variety of phenomena [2], which span from the nanoscale confinement of the superconducting condensate [3] and the emergence of unconventional (equal-spin triplet) pairing [4,5] to the tuning of magnetization dynamics by superconductivity [6,7]. One of the topics receiving much continued attention is the pinning and manipulation of Abrikosov vortices (flux quanta) by ferromagnetic "templates" [8,9], such as patterned arrays of nanomagnets [10–12] or the domain structure of ferromagnetic films [13–16]. The existing research focuses both on fundamental and technological aspects [17], since controlling the dynamics of flux quanta allows the design of the magneto-transport properties of superconductors. For instance, the magnetic pinning of Abrikosov vortices can increase the critical current (and thus decrease the electrical resistance and noise) in superconducting devices. Moreover the control of vortex motion opens the door to novel concepts of superconducting electronics and computing [18–22], from electrical rectification [18] to quantum cellular automata [19].

In this context, superconductor/ferromagnet (S/F) multilayers –in which the vortex manipulation in the S is achieved through their interaction with the stray magnetic field from the F domains [13–16] – are specially interesting. This is because the domain structure can be reconfigured by the history of applied magnetic fields, which allows setting different domain distributions (and thus different vortex pinning landscapes) in a single device [13–16]. Consequently, the superconducting transport of a S/F multilayer may be strongly dependent on the magnetic history – a property absent in bare superconductors. Being able to tailor this behavior is one of the keys to obtain functionality from it.

In this article, we show that the field range in which the above memory effects develop and their strength can be designed by engineering the size and morphology of the magnetic domain texture. This is realized with multilayers composed of ultrathin layers of a ferromagnet (Co) intercalated between heavy metal layers (Ir, Pt, Ru). By choosing the thickness and stacking order of the layers, we can design the magnetization reversal so that, within a similar range of applied fields, the ferromagnetic multilayer shows either (i) perpendicularly magnetized domains of variable size or even (ii) magnetic skyrmions [23,24]. These latter are sub-micrometric magnetization textures with non-trivial topology [see scheme in Fig. 1 (g)], which may be distributed regularly [see Fig. 1 (f)] and whose interaction with Abrikosov vortices has been recently explored theoretically [25–28]. We compare vortex pinning by the different types of magnetic textures (i) and (ii) using both magneto-resistance and critical current measurements, which are analyzed using a specifically developed analytical model. This allows calculating the vortex pinning enhancement due to the Meissner currents induced by the stray field from the ferromagnet. In other terms, we consider only magnetic interactions, which favor vortex localization wherever the stray field has same magnetic polarity and is the strongest so as to minimize the magnetic energy. Note, however, that we neglect the depression of superconductivity due to the stray fields (these are much lower than the upper critical field $H_{c2}$), as well as the proximity effect (ruled out because the pinning effects appear in the presence an insulating alumina interlayer between the superconductor and the ferromagnet).

We find that the ability to tune the characteristic domain size allows tailoring magnetic vortex pinning via two effects. First, the strength of the stray magnetic field depends on the domain size, which therefore affects the pinning energy. Second, through a geometrical effect: when the (field dependent) inter-vortex distance is comparable to the average domain size, a greater number of vortices sit near domain walls, where the stray magnetic field is the strongest. This increases the overall pinning within the superconducting film, similarly as observed with periodic pinning arrays [8]. The relative contribution of those two effects determines the field range over which the magnetic vortex pinning is the strongest. Finally, we show that although the presence of skyrmions results in magnetic pinning, it is lower than expected for stripe domains of comparable width. As we discuss below, this is because skyrmions behave as point-like obstacles for vortices, thus impeding vortex-motion less efficiently than elongated domains.

## II. METHODS

Samples were fabricated by dc sputtering at room temperature in an argon atmosphere. First, a 60 nm-thick film of the amorphous superconductor $Mo_{80}Si_{20}$ (thereafter denoted as MoSi) with $T_c \approx 6.5$ K is deposited on an undoped silicon substrate. This sample is then cut in different pieces, some of which are spared as reference samples. Finally, different magnetic multilayers are deposited *ex situ* on the MoSi using the same sputtering technique. We studied three multilayers: $AlO_{x\ 3nm}/Pt_{10nm}/(Co_{0.6nm}/Pt_{1nm})_5/Pt_{3nm}$ (thereafter referred to as Co/Pt), $AlO_{x\ 3nm}/Pt_{10nm}/(Ir_{1nm}/Co_{0.6nm}/Pt_{1nm})_5/Pt_{3nm}$ (thereafter Ir/Co/Pt), and $Ta_{5nm}/Pt_{8nm}/(Pt_{1.2nm}/Co_{1.6nm}/Ru_{1.4nm})_4/Pt_{3nm}$ (thereafter Ru/Co/Pt). The single S films and S/F structures were patterned into a multi-probe transport bridge (200 µm-long and 40 µm-wide) that allows measuring the longitudinal and transverse resistivities. The patterning is done using conventional UV lithography and argon ion milling. Four-probe magneto-transport measurements were carried out either in a liquid-He cryostat or in a close-cycle refrigerator, both equipped with an electromagnet producing a magnetic field perpendicular to the film plane. The dc resistance $R = V/I$ was measured by injecting an electrical current $I$ with a dc source and measuring the voltage $V$ with a voltmeter. The voltage offsets are removed by measuring both current polarities. The critical current $J_c$ is calculated from V(I) curves, using a voltage criterion of 5 µV ($E_C = 25$mV.m$^{-1}$). The magnetic domain structure is characterized by room temperature

Magnetic Force Microscopy (MFM) in 'lift mode', at a typical height of 30 nm above the surface.

## III. EXPERIMENTAL RESULTS

Figure 1 shows room temperature MFM images of the domain structures recorded for samples based on the three different ferromagnetic multilayers: Co/Pt, Ir/Co/Pt and Ru/Co/Pt.

Co/Pt multilayers constitute a well-known perpendicular magnetic anisotropy system whose demagnetized state (mean out-of-plane magnetization $m_z \equiv M_z/M_S = 0$) can present a maze-like distribution of opposite, perpendicularly magnetized domains, as shown in Fig 1 (a). The characteristic domain width is $L/2$ ~ 470 nm, with $L$ the characteristic periodicity of the domain structure, obtained from an autocorrelation analysis of the MFM image. Upon application of a magnetic field perpendicular to the film, the parallel (antiparallel) domains gradually grow (shrink), thus increasing the mean magnetization [see image for $m_z = 0.5$ in Fig. 1 (b)]. The intercalation of an additional heavy element layer (Ir or Ru) leads to a reduction of the characteristic domain period $L$ observed at $m_z = 0$ [see Fig. 1 (c) and (e), from which we obtain $L \approx 220$ nm and $\approx 170$ nm respectively], and finally to the stabilization of magnetic skyrmions upon application of a magnetizing field [see Fig. 1 (d) and (f) for $m_z = 0.5$]. The formation of skyrmions as well as the periodicity reduction are favored by the Dzyaloshinskii-Moriya Interaction (DMI), which arises due to symmetry breaking at the interfaces and the strong spin-orbit coupling in the heavy metal [23,24]. Note that the symmetric stacking Pt/Co/Pt should result in no net DMI, because the contributions from Pt/Co and Co/Pt interfaces cancel each other. Without DMI, the domain walls cost energy and the presence of magnetic domains is only favored by the dipolar energy reduction. On the contrary, in the asymmetric stacks Ru/Co/Pt and Ir/Co/Pt, a net effective interfacial DMI exists (of the order of 1 pJ/m). The interfacial DMI favors a non-collinear (cycloidal-like) arrangement of the magnetization in space that it is in competition with ferromagnetic exchange (a few tens of pJ/m) and the interfacial perpendicular anisotropy (which both favor parallel alignment of the magnetization). This competition results in reduced domain wall energies, and hence smaller magnetic domains (See Appendix A). The DMI also fixes the chirality of the spin textures through its sign and, if properly balanced with all the other magnetic interactions at play, leads to the formation of Néel skyrmions [23,24,29,30] [see scheme in Fig. 1 (g)]. Note that skyrmions appear scattered and mixed with elongated domains of the same polarity in the Ir/Co/Pt multilayer [Fig. 1 (d)], while they distribute more regularly in the Ru/Co/Pt multilayer [Fig. 1 (f)] due to lower effective perpendicular anisotropy [23].

Fig. 2 (a)-(c) show magnetization loops $m_z(H)$ as deduced from Hall effect measurements at 10 K (*i.e.* above the superconducting critical temperature $T_c$) with $H$ perpendicular to the film plane (along the $z$-axis). In particular, we assume that the Hall resistivity is dominated by the Anomalous Hall Effect as shown for similar multilayers [31], and is therefore proportional to the reduced magnetization along the $z$-axis, $m_z(H)$ [32]. In addition to the field increasing (red) and decreasing (black) branches, i.e. those corresponding to the reversal from negative to positive magnetic saturation and *vice versa*, a third branch is shown (blue) which starts from a demagnetized state at $H = 0$. The samples are demagnetized by cycling the field between positive/negative polarities, starting from saturation and reducing the amplitude by 10% each iteration.

Both Co/Pt [Fig. 2 (a)] and Ir/Co/Pt [Fig 2. (b)] show 100% remanence, but the former multilayer has a much sharper magnetization reversal and a smaller coercive field, as Co/Pt multilayers have a larger perpendicular anisotropy than Ir/Co/Pt. In both samples, the demagnetized state is quite stable against the applied field [see blue curves in Fig. 2 (a) and (b)]. On the other hand, the Ru/Co/Pt sample [Fig. 2 (c)] was designed to have a nearly vanishing out-of-plane magnetic anisotropy, orders of magnitude smaller than the two other samples, and shows a radically different behavior. First, the magnetization abruptly decreases before the applied field is actually reversed. Second, it shows nearly no remanence. Third, the curve measured after demagnetization (blue) falls on top of the field-increasing branch (red).

Magnetoresistance $R(H)$ measurements below $T_c$ ($T$ = 3.5 K), for the same field sweeps (and same color code) as for the magnetization curves in (a)-(c) are plotted in Fig. 2 (d)-(f). In these measurements, the magnetoresistance is strongly hysteretic for all the S/F samples, which is in contrast with the behavior of bare MoSi films [green curve in each panel of Fig. 2 (d)-(f)].

In the case of Co/Pt [Fig. 2 (d)], sharp dips around the positive/negative coercive fields are observed. The resistance is lower during the magnetization reversal than when the sample magnetization is saturated. Notably, the lowest resistance is reached after the sample has been demagnetized [blue curve in Fig. 2 (d)]. When comparing the S/F magneto-resistance with that of the single MoSi film (green curve), two different field regimes have to be distinguished. For a field large enough to saturate the F magnetization, the S/F multilayer and the single S film show similar magnetoresistances. At intermediate and low fields, the S/F magnetoresistance is lower than that of the MoSi film *only* in the presence of magnetic domains, *i.e.* across the magnetization reversal or when the sample has been demagnetized (blue curve). Contrarily, if the magnetization is saturated ($|m_z| = 1$) and below $\mu_0 H \approx$ 60 mT, the resistance becomes significantly higher for the S/F multilayer than for the single S film.

The Ir/Co/Pt sample [Fig, 2 (e)] behaves qualitatively similar to the Co/Pt one. The main difference is that the resistance drop around the coercive fields extends over a larger field range, as the magnetization reversal does [see Fig. 2 (b)]. In the case of Ru/Co/Pt [Fig. 2 (f)], the resistance drop associated with the magnetization reversal (presence of magnetic domains) extends over an even broader field range, following the magnetization curve [Fig. 2 (c)]. Furthermore, the resistance drop is much stronger than in the two previous samples. Consider for instance the resistance around $\mu_0 H \approx$ 50 mT (red curve) and -50 mT (black curve), which is in the $10^{-3}$ Ω range for this sample [Fig. 2 (f)] and one order of magnitude higher for the two others [Fig. 2 (d) and (e)]. Notice also that, in the field regime $\mu_0|H|$< 30 mT and regardless of the magnetic history (red, black and blue curves), the Ru/Co/Pt sample's magnetoresistance is lower than that of the single S film (green curve), which is in stark contrast with the two other S/F samples. This is because, regardless of the field history, the Ru/Co/Pt magnetization is broken into domains ($|m_z| < 1$) for $\mu_0|H| \lesssim$ 30 mT.

The effect of the different domain structures can be quantified by considering the field-dependent critical current of the different S/F multilayers, $J_{C,SF}(H)$ and that of the bare MoSi control samples, $J_{C,S}(H)$ [see examples of $J_C(H)$ in the insets of Fig. 3 (d)-(f)]. Indeed, a direct signature of the critical current variation due to the presence of F can be obtained by considering $\Delta J_C(H) = J_{C,SF}(H) - J_{C,S}(H)$, as shown in Fig. 3 (d)-(f). The three colored curves in each figure correspond to the same three different magnetic field sweeps (same color code) followed in the magnetization loops [Fig. 3 (a)-(b)]. The behavior is consistent with that observed in the magnetoresistance, and presents a series of features common for all samples. A net critical current enhancement ($\Delta J_C > 0$) is observed when the F is broken into domains ($|m_z| < 1$) – that is, during magnetization reversal and after demagnetization (blue). When the F is uniformly magnetized ($|m_z| = 1$), two regimes can be distinguished. When the magnetic fields are low, $\Delta J_C < 0$ which means that the presence of the F leads to a decrease of the critical current. At high fields $\Delta J_C \approx 0$ , which proves that in that regime, the presence of F does not have a strong effect on $J_C$. Besides these common features, there are strong differences from sample to sample. Qualitatively, we can see that the shape of curves measured after demagnetization (blue) are very different for the three samples. Also, while there is a qualitative resemblance between Co/Pt [Fig. 3 (d)] and Ir/Co/Pt [Fig. 3(e)] concerning the sweeps from negative and positive saturation (red and black curves), in the case of Ru/Co/Pt the curve is very different [Fig. 3(f)]. Quantitatively, we find that the maximum critical current enhancement is around twice lower for the Ir/Co/Pt [Fig. 3 (e)] than for the two other samples. However, it should be noticed that for Co/Pt [Fig. 3 (d)] the maximum critical current enhancement is observed after demagnetization (blue curve), while for Ru/Co/Pt [Fig. 3(f)] it is observed at for $H$=0 regardless of the magnetic history.

## IV. THEORETICAL MODELLING

Based on the stacking of the samples, which is detailed in Figure 4 (a), we construct a simple model to capture how the varying domain structures might enhance the critical current. This model is sketched in Fig. 4 (b). We consider a superconducting film of thickness $d_s$. The magnetic multilayer, which consists of N repetitions of Co/heavy element layers (Co/Pt, Ir/Co/Pt or Ru/Co/Pt), is modeled as a single ferromagnetic film of saturation magnetization $M_s$ and thickness $d_f = N \times d_{Co}$, with $d_{Co}$ the thickness of the individual Co layers in the structure. The ferromagnetic and superconductor films are separated by a distance $a$, which accounts for the presence of buffer layers in the structure. In the ferromagnet, the magnetization points out-of-plane (along the $z$ axis). For simplicity, we consider a periodic domain structure consisting of stripes parallel to the $y$-axis. Their width is $w_+ = w$ for "up" domains, and $w_- = L - w$ for "down" ones, with $L$ being the periodicity of the structure. The ratio $w_+/w_-$ is fixed by the external field, while $L$ depends only on the material properties. Considering that the ferromagnet is thin ($d_f << L$) and that the width of the domain walls is negligible compared to $L$, we approximate the magnetization profile as an alternating step-like function along $x$. The stray magnetic field $B_F$ from such a domain structure can be analytically calculated (see Appendix B) and has components only along $x$ and $z$. Since the superconducting film is thin with respect to the magnetic penetration depth ($d_s = 60$ nm $\ll \lambda \approx 270$ nm as estimated from the normal state resistivity and $T_c$ measurements [33]), we consider only the component along $z$. At the surface of the superconductor ($z = 0$), that component reads:

$$B_{Fz}(x) = \frac{4\mu_0 M_S d_f}{L} \sum_{n\geq 1} \sin\left(\frac{\pi n w}{L}\right) \cos\left(\frac{2\pi n x}{L}\right) e^{-2\pi n a/L} \quad \text{(Eq. 1)}$$

Note that the magnetic field screening in a thin superconducting film is determined by Pearl's length $\Lambda = \lambda^2/d_S \approx 1200$ nm, instead of $\lambda$ as in bulk materials [34]. Thus, the induced Meissner current density in the superconductor reads:

$$j_m(x) = -\frac{2M_S d_f}{\pi \Lambda d_s} \sum_{n\geq 1} \frac{1}{n} \sin\left(\frac{\pi n w}{L}\right) \sin\left(\frac{2\pi n x}{L}\right) e^{-2\pi n a/L} \quad \text{(Eq. 2)}$$

$B_{F,z}(x)$ and the corresponding $j_m(x)$ for the domain structure sketched in Fig. 4 (b) are respectively shown in Fig. 4 (c) and (d). We find that the induced Meissner current is the highest near the domain walls and cancels in the center of the domains. The vortex depinning current can thus be estimated from the current that equals the maximum of the Meissner current,

$$j_p = max|j_m| \quad \text{(Eq. 3)}$$

To calculate $\Delta J_C(H)$ using this model, we must account for the field-induced evolution of the magnetic domain widths $(w_+, w_-)$. The magnetization increases under an applied magnetic field $H_z$ because domains with the polarity along the applied field $H$ grow ($w_+$ increases) while those with opposite polarity shrink ($w_-$ decreases). Thus, the reduced magnetization is given by $m_z = (w_+ - w_-)/L = 2w/L - 1$, and the domain width evolves as a function of the applied field following the magnetization :

$$w(H) = \frac{L(m_z(H)+1)}{2} \quad \text{(Eq. 4)}$$

For each sample, we calculate $\Delta J_C(H)$ from Eqs. 2, 3 and 4 using : i) $L$ as estimated from the MFM images at $m_z = 0$ (Fig. 1); ii) $m_z(H_z)$ shown in Fig. 2 (a)-(b) for the different field sweeps; and iii) $M_S$ = 1.09 $MA.m^{-1}$ for Co/Pt, $M_S$ = 0.9 $MA.m^{-1}$ for Ir/Co/Pt, and $M_S$ = 1.21 $MA.m^{-1}$ for Ru/Co/Pt, as obtained from magnetization measurements at 300 K. The results are shown in Fig. 3 (g)-(h)-(i) for comparison with the experimental data. Note that our model makes several assumptions. First, the magnetic texture is simplified to a distribution of parallel stripe domains, and the domain wall width is neglected. Second, the critical current is estimated as the maximum of the Meissner current. While both must be proportional, their ratio between is not necessarily equal to one as assumed. Finally, temperature effects are considered only through the penetration depth $\Lambda$. As we discuss in the next section, these approximations impact the results at the quantitative level. However, the model allows for a good qualitative understanding of the field dependence of the critical current in the presence of magnetic domains.

## V. DISCUSSION

As we detail below, the hysteretic critical current enhancement calculated with the above model [Figs. 3 (g)-(h)-(i)] reproduces well the main features of the experimental $\Delta J_C(H)$ curves [Fig. 3 (d)-(e)-(f)], and therefore explains the hysteretic magneto-resistance observed in Fig. 2 (d)-(e)-(f). This implies that the pinning enhancement is produced by the Meissner current $j_m$ induced by the stray magnetic field from the domain structure. The Meissner exerts a Lorentz force $F_L \propto j_m \phi_0$ on the vortices. This is maximum at the domain walls, cancels at the center of the domains, and changes sign following the distribution of up/down magnetized domains (Fig. 4). Consequently, vortices are pushed by the resulting force towards the center of the domains of the same polarity. This illustrates how the Meissner current distribution creates an energy landscape with minima in which vortices are pinned. The model calculations show that the maximum Meissner current (or maximum Lorentz force), which yields the depth of the minima in the energy landscape, depends on the domain sizes. Therefore, it evolves with the applied magnetic field as the domain structure does, resulting in a hysteretic behavior. In addition to that main mechanism, three effects that are not considered within the model can explain the incidental discrepancies between theory and experiments. These are vortices induced by the ferromagnet, commensurability between the vortex lattice and the domain structure, and field-induced changes in the domain morphology.

In order to support the above and to highlight the role of the different effects, we compare point-by-point experiments and theory for each sample in what follows.

We will start with the case of Ir/Co/Pt, for which the calculated $\Delta J_C(H)$ qualitatively reproduces the experimental one very closely, see Fig. 3 (e) and (h). We see that the curve measured after demagnetization (blue) presents a plateau up to $\mu_0 H \approx 75$ mT, which is also found in the model, and corresponds to the field range in which the magnetization is constant ($m_z \approx 0$). Thus the domain structure remains unchanged, and consequently so does the pinning landscape. When the applied field is increased above $\mu_0 H \approx 75$ mT, the domains parallel to the applied fields grow and $m_z$ increases, leading to a gradual decrease in the magnetic pinning. Once the magnetization is fully saturated ($m_z = 1$), the stray magnetic field $B_{Fz}(x) \approx 0$, and therefore the pinning landscape becomes flat, leading to $\Delta J_C(H) \approx 0$. The model also captures the increase of $\Delta J_C(H)$ across the magnetic reversal (red curve). In this branch, $\Delta J_C(H)$ increases until a balanced distribution of domains is obtained ( $m_z \approx 0$ at $\mu_0 H \approx 80$ mT), which yields a rough pinning landscape. Further increase of the magnetic field yields to a steady decrease of $\Delta J_C(H)$ as the average $m_z$ increases, due to domain expansion and the subsequent flattening of the pinning landscape. The only discrepancy between the experiment and the model occurs indeed at low fields when $|m_z| = 1$. In this limit, the experiments show a net decrease in the critical current ($\Delta J_C(H) < 0$) that is not reproduced by the model. This disagreement can be explained by the nucleation of vortices caused by the magnetized F. Notice that the model considers the pinning of vortices induced by the external field, which yields a critical current increase (similarly a decrease of magneto-resistance). However, it does not take into account the possible additional generation of vortices caused by the presence of the F, which has the opposite effect. The experiments show that the latter effect is dominant at low applied fields (when there are few vortices induced by the external field) if the F is magnetized ($|m_z| = 1$). This is because the magnetized F increases the vortex population through two mechanisms. On one hand, via finite size effects: near the edges of sample (lithographed transport bridge), fringe fields lead to the local nucleation of vortices, which can enter the bulk of the film where the magnetic pinning landscape is flat. Furthermore, if we consider a fully magnetized F and the magnetic field from vortices, we find that vortex entrance lowers the system's magnetostatic energy [34]. These two mechanisms increase the vortex population beyond that induced by the external field. This additional population is significant at low external fields, thereby producing a decrease of the critical current ($\Delta J_C(H)$<0), but gradually becomes negligible – and consequently its effects on $\Delta J_C(H)$ and $R(H)$ too – as more vortices induced by the internal field penetrate the S.

For the case of Co/Pt, experiments [Fig. 3 (d)] and theory [Fig 3 (g)] agree similarly as for Ir/Co/Pt only for the fields sweeps that start from magnetic saturation (red and black curves). However, in the measurement made after demagnetization (blue curve), a strong peak develops [Fig. 3 (d)] which contrasts with the plateau theoretically expected [Fig 3 (g)] for the constant $m_z \approx 0$ [Fig. 3 (a)]. This peak appears around $\mu_0 H_z$≈10 mT, which corresponds to a distance between vortices $d \approx \sqrt{\phi_0/\mu_0 H} \approx$ 450 nm that matches the average domain width $w \approx$ 470 nm [see discussion on Fig. 1 (a)]. This suggests that the peak results from commensurability between the vortex-lattice and the domain structure. Indeed, this effect is not observed in the two other systems because the magnetic domains are much smaller. In particular, $w \approx$ 110 for Ir/Co/Pt nm and $w \approx$ 85 nm for Pt/Co/Ru as measured by MFM. For such domain sizes , the matching fields would be respectively of 170 and 300 mT, which are beyond the saturation fields of those multilayers.

In the case of Ru/Co/Pt, the model explains the strong critical current enhancement observed at low fields, which is associated to the presence of domains ($m_z \approx 0$). It also accounts for the decay of $\Delta J_C(H)$ with increasing $H$. Nevertheless, in the experiments the decay is faster than expected from the simulations. This is because the model does not consider changes in the domain morphology. Yet in this sample, a regular array of skyrmions is rapidly formed upon increasing $H$ [Fig. 1 (f)]. Thus, unlike in Co/Pt and Ir/Co/Pt, for which elongated domains and $m_z \approx 0$ persist over a large field range, for Ru/Co/Pt the model of stripe domains becomes rapidly a poor approximation as $H$ (and $m_z$) increases. The main reason is that, due to their relatively small size and morphology, skyrmions (which repel vortices due to their opposite magnetic polarity [26]) constitute point-like obstacles for vortices. Thus, vortices can more easily slip around skyrmions than across elongated magnetic domains when pushed by the Lorentz force.

Finally, if we compare the values of the critical current enhancement for low fields and $m_z \approx 0$, we see that the model approximately predicts the hierarchy between the different samples. In particular, if we discard matching effects in Co/Pt, we observe that the strongest critical current increase corresponds to Ru/Co/Pt, as expected from the model. To a large extent, this is because it has the highest magnetic moment per unit area $M_s\, d_f$ . Thus, even if the domain size is comparable to that in Ir/Co/Pt, Ru/Co/Pt presents a much higher critical current enhancement. Conversely, Co/Pt has a similar magnetic moment to that of Ir/Co/Pt, but provides a slightly higher enhancement due to the domains being larger. Note that for all samples, the predicted $\Delta J_C(H)$ is up to 50 to 100 times higher than the measured one. Two reasons might explain this discrepancy. First, the model assumes vortex depinning perpendicular to the stripes. However, in the studied samples the elongated domains are disordered and point along many different directions with respect to the injected current (depinning Lorentz force). As shown earlier [35], stripe domains not perpendicular to the Lorentz force pin vortices much less effectively, since vortices can glide along the domain walls, and therefore provide a much weaker contribution to the critical current enhancement. Thus, the disordered domains expectedly yield a critical current enhancement much lower than predicted by the parallel stripes model. Second, the model considers temperature effects only through the temperature dependence of $\Lambda$, but thermal depinning effects − which may strongly attenuate the critical current enhancement [36] − have not been included in it.

## VI. CONCLUSION

We have studied Abrikosov vortex pinning in F/S hybrids made of superconducting thin films covered by ferromagnetic multilayers based on Co and different heavy metals. The ferromagnetic domain morphology and the magnetization reversal is engineered via changes of the multilayers stacking, which controls the balance between perpendicular magnetic anisotropy and Dzyaloshinskii–Moriya interactions. A crossover from large and elongated domains (hundreds of nm) to small magnetic skyrmions (tens of nm) can be obtained in this manner, which drastically changes the vortex pinning properties and, accordingly, the critical current. Qualitatively, the magnetic pinning is well described by a model that explains the critical current enhancement based on the magnitude of the stray magnetic field generated by the domain structure. An additional effect, geometric matching between the domain structure and the vortex distribution can further enhance pinning effects. We also show that the presence of magnetic domains enhances the critical current, whereas a homogeneous magnetization tends to degrade it. This drawback can be overcome by engineering the F magnetization reversal so that it shows nearly no remanence. We also found that a regular distribution of skyrmions produces lower pinning than expected for stripe domains of comparable width. This is essentially because the skyrmions are point-like objects, allowing the vortices to circumvent them. We also anticipate much stronger pinning due to commensurability effects if skyrmions could be stabilized in a field range in which they have the same polarity as vortices. Altogether, these results provide a key insight on how to tailor the vortex pinning in S/F bilayers through magnetic domain engineering, which allows optimizing the pinning enhancement in the desired field ranges.

**Acknowledgements**

Work supported by the ERC grant Nº 647100 “SUSPINTRONICS”, French ANR grants ANR-15-CE24-0008-01 “SUPERTRONICS” and ANR-17-CE30-0018-04 “OPTOFLUXONICS”, and European COST action 16218 “Nanocohybri”. J. S. thanks INP-CNRS and "Scholarship program Alembert" funded by the IDEX Paris-Saclay ANR-11-IDEX-0003-02 for support during his stay at the Unité Mixte de Physique CNRS/Thales. A.V.S. acknowledges the funding from by Russian Science Foundation under Grant No. 17-12-01383 (Sec. IV). We thank F. Ajejas for support during the magnetization measurements and W. Legrand for helpful discussions.

## APPENDIX A: ROLE OF DMI FOR MAGNETIC DOMAIN SIZES.

The size of magnetic domains results from the competition between different energies: dipolar energy, symmetric (Heisenberg) and antisymmetric (DMI) exchange, anisotropies, and external fields. The presence of domains reduces the dipolar energy, but the associated domain walls (DW) cost an energy density given by $\sigma_{DW} = 4\sqrt{AK} - \pi D$, with $A$ the exchange stiffness, $K$ the effective wall anisotropy and $D$ the effective DMI [37–39]. The calculation of the domain size for magnetic multilayers is complex but has been done for simplified models with uniform magnetization through the multilayer thickness [40], as well as for the complex case where the DMI competes with the stray field of the domains, resulting in hybrid textures [41,42].

In our study, we are only interested in the stray field from the multilayer, and we considered the approximation of a DW of zero width to have a handleable model. So DMI plays a role in fixing the period of the domains, but the internal DW texture is neglected. The period of the domains at zero external field is measured experimentally by MFM.

**APPENDIX B: DETAILS ON THE THEORETICAL MODEL.**

We consider a superconductor-ferromagnet bilayer as represented in Fig. 4(a). We assume that the ferromagnetic layer is thin ($d_F \to 0$). Its domain structure consist of stripes parallel to the $y$ axis, with out-of-plane magnetization (along $z$). The up domains have width $w$, and those opposite $L-w$, where $L$ is the structure's periodicity. If the domain-wall width is much smalled than $w$, then a good approximation of the magnetization is a step-like function along the $x$-axis:

$$\mathbf{M} = M(x)\mathbf{e_z} \quad ; \quad M(x) = \begin{cases} M_s, & -w/2 + Ln \leq x \leq w/2 + Ln \\ -M_s, & w/2 + Ln \leq x \leq -w/2 + L(n+1) \end{cases} \tag{1}$$

where $\mathbf{e_x}, \mathbf{e_y}, \mathbf{e_z}$ are the unitary vectors of the Cartesian frame and $n$ is an integer. The Fourier expansion of the magnetization then writes :

$$M(x) = \sum_{n=0}^{\infty} M_n e^{iqnx} \quad ; \quad M_n = \frac{1}{L}\int_{-L/2}^{L/2} M(x)e^{-iqnx}dx \quad ; \quad q = \frac{2\pi}{L} \tag{2}$$

$$M_{n=0} = M_S\left(\frac{2w}{L} - 1\right); \quad M_{n\neq 0} = \frac{2M_S}{\pi n}\sin\left(\frac{\pi n w}{L}\right) \tag{3}$$

Consequently, the magnetization of the F layer can be represented as :

$$\mathbf{M} = M_S\mathbf{e_z}\delta(z+a)\left\{\frac{2w}{L} - 1 + \frac{2}{\pi}\sum_{n\neq 0}\frac{1}{n}\sin\left(\frac{\pi n w}{L}\right)e^{2i\pi n x/L}\right\} \tag{4}$$

$$\delta(z+a) = \int \frac{dp}{2\pi} e^{ip(z+a)}. \tag{5}$$

Following that, the magnetic current sheet density $j_F = d_F \nabla \times \mathbf{M}$ only has a component along $y$, which is :

$$\mathbf{j_F} = -d_F\mathbf{e_y}\frac{\partial M}{\partial x}\delta(\mathrm{z}+\mathrm{a}) =$$

$$-\frac{4iM_S d_f}{L}\mathbf{e_y}\int \frac{dp}{2\pi}e^{ip(z+a)}\sum_{n\neq 0}\sin\left(\frac{\pi n w}{L}\right)e^{i2\pi n x/L} \tag{6}$$

As a result, the vector potential $\mathbf{A_F} = A_F(x,z)\mathbf{e_y}$ due to the ferromagnet is also directed along $y$ and is obtained from the Maxwell equation :

$$\nabla \times \nabla \times \mathbf{A_F} = -\nabla^2 \mathbf{A_F} = \mu_0 \mathbf{j_F} \tag{7}$$

$$A_F = \int \frac{dp}{2\pi} e^{ipz} \sum_{n\neq 0} A_{Fn} e^{iqnx} \quad ; \quad A_{Fn} = -\frac{2i\mu_0 M_s d_f}{\pi L} \sin\left(\frac{\pi n w}{L}\right) \frac{e^{ipa}}{p^2 - q^2 n^2} \tag{8}$$

$$A_F(x,z) = \frac{2\mu_0 M_s d_f}{\pi} \sum_{n\geq 1} \frac{1}{n} \sin\left(\frac{\pi n w}{L}\right) \sin\left(\frac{2\pi n x}{L}\right) e^{-2\pi n(z+a)/L} \tag{9}$$

In experiment, we have $L \ll \Lambda = \lambda^2/d_s$. We may then neglect the superconducting screening to calculate $A_F$.

The stray magnetic field $B_F$ from the domain structure only has components along $x$ and $z$. As the superconductor is thin ($d_s \ll \lambda$), we consider only the latter. At the surface of the superconductor ($z = 0$), it writes :

$$B_{Fz}(x) = \frac{4\mu_0 M_s d_f}{L} \sum_{n\geq 1} \sin\left(\frac{\pi n w}{L}\right) \cos\left(\frac{2\pi n x}{L}\right) e^{-2\pi n a/L} \tag{11}$$

The resulting Meissner sheet current density is then given by :

$$j_m(x) d_s = -\frac{1}{\mu_0 \Lambda} A_F(x, z=0) =$$

$$-\frac{2 M_s d_f}{\pi \Lambda} \sum_{n\geq 1} \frac{1}{n} \sin\left(\frac{\pi n w}{L}\right) \sin\left(\frac{2\pi n x}{L}\right) e^{-2\pi n a/L} \tag{12}$$

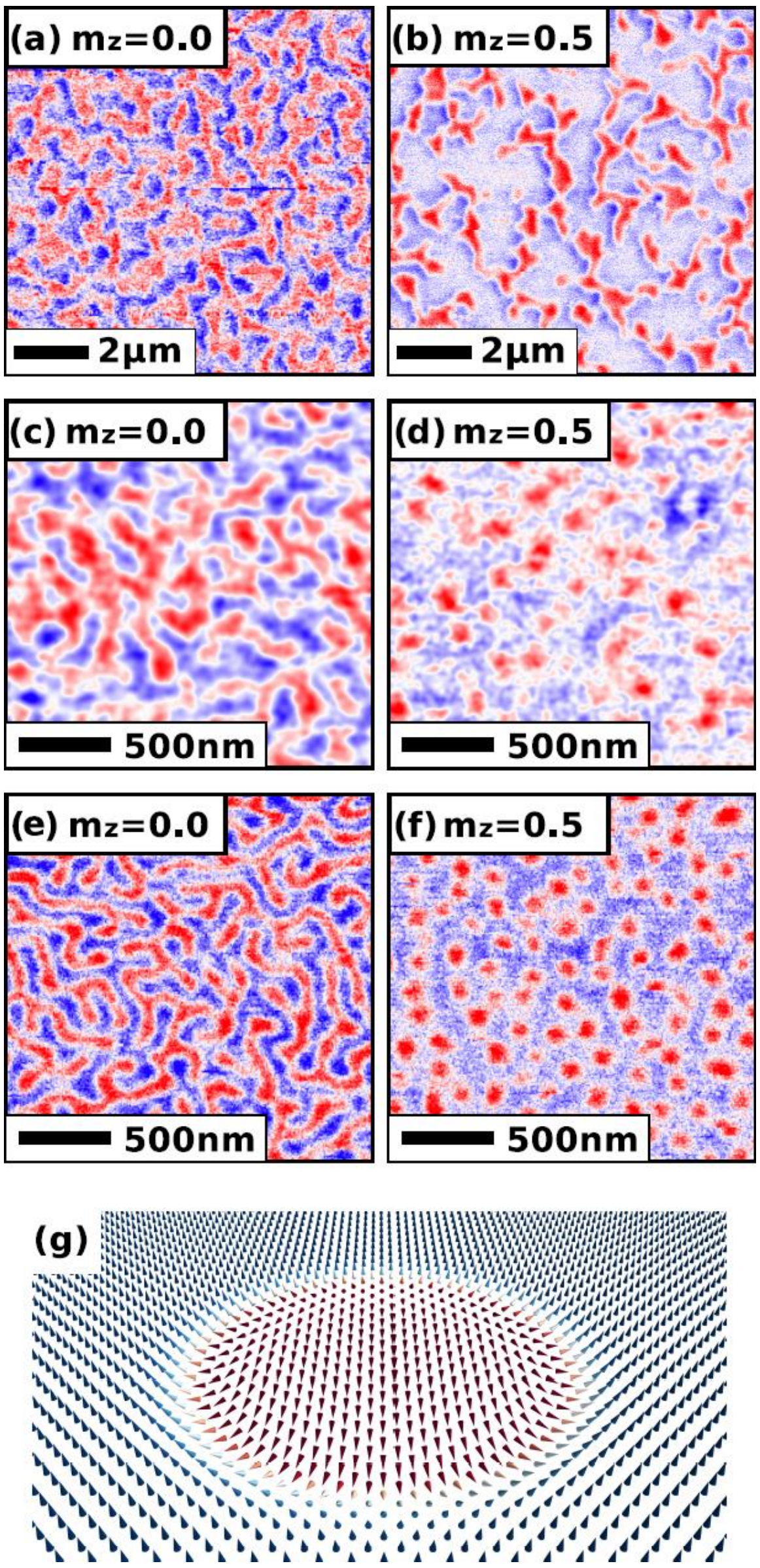

**Figure 1:** MFM images at 300K for Co/Pt (a)-(b), Ir/Co/Pt (c)-(d), and Ru/Co/Pt (e)-(f) multilayers. For each sample, a series of images was acquired in increasing magnetic fields, to monitor the magntization reversal from negative ($m_z = -1$) to positive ($m_z = 1$) saturation. Images are shown for $m_z = 0$ and $m_z = 0.5$, as estimated from the up/down domain areas. (g) Sketch of a Néel skyrmion, in which cones represent the orientation of the local magnetic moments. The color code highlights the direction of $m_z$, correspondingly to the MFM images. In images (d) and (f), the skyrmions appear as the red disks.

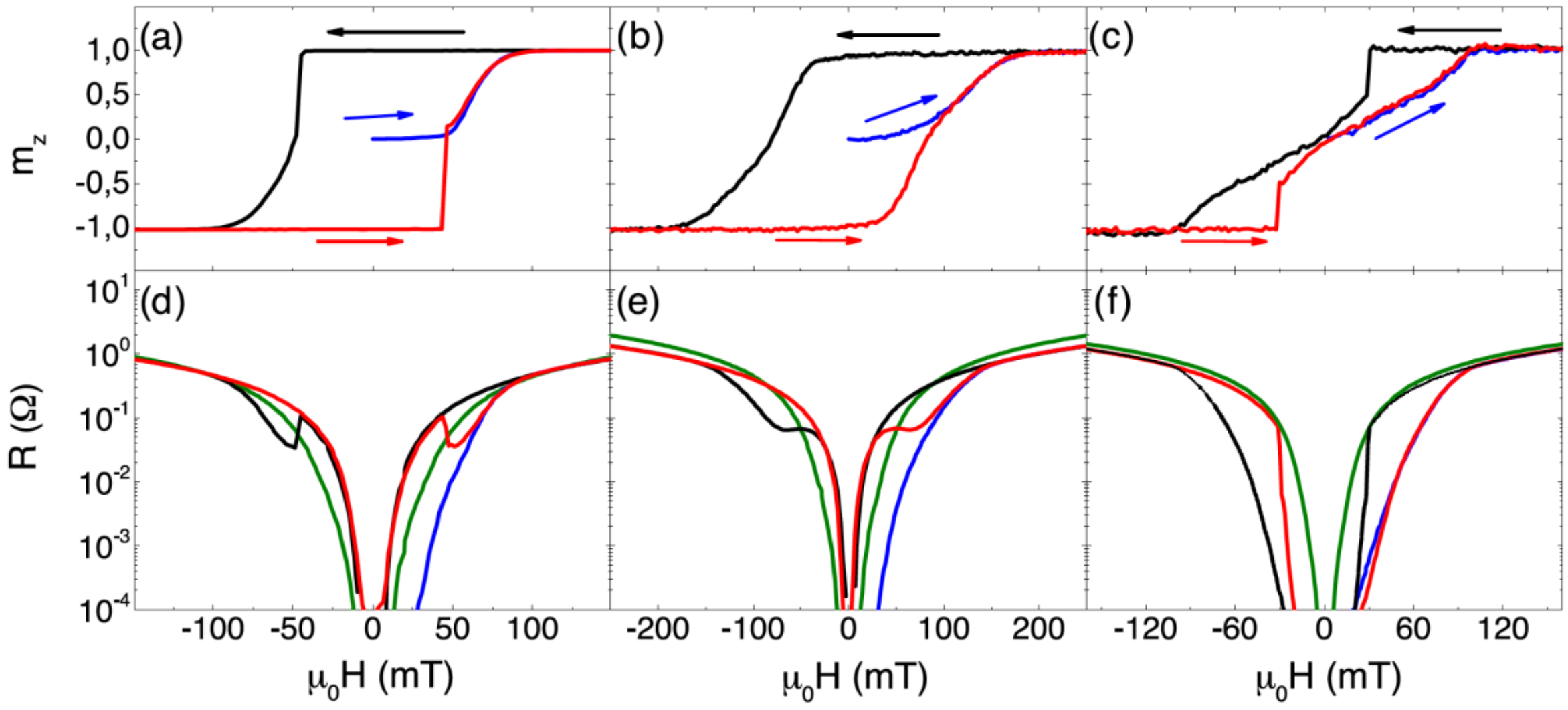

**Figure 2:** (a)-(b)-(c) magnetization loops $m_z(H) = M_z(H)/M_S$ deduced from Hall effect measurements in the normal state ($T$=10 K ; $I$=500 µA), respectively for Co/Pt, Ir/Co/Pt and Ru/Co/Pt samples. The arrows and color code indicate the magnetic field sweep sequence. (d)-(e)-(f) show the longitudinal magnetoresistance loops in the superconducting state ($T$=3.5 K ; $I$=1mA). The color code (black, red, blue) refers to the field sweeps indicated in (a)-(b)-(c). Green curves are measurements made in the same conditions with the plain MoSi films used as reference.

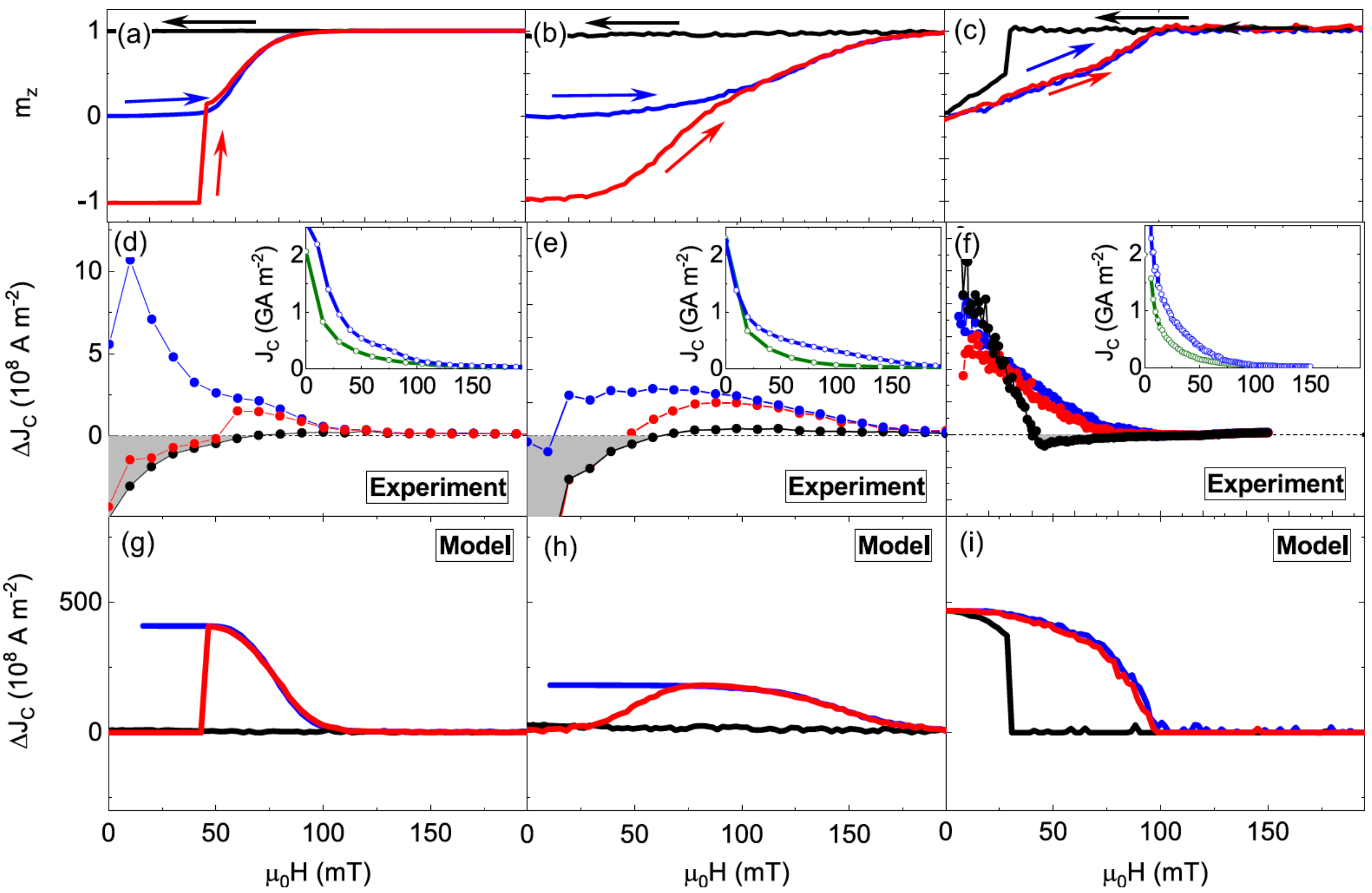

**Figure 3:** (a)-(b)-(c) Zoom for positive fields of the magnetization loops shown in Fig. 2 (a)-(b)-(c), for Co/Pt, Ir/Co/Pt, and Ru/Co/Pt respectively. The magnetic field is swept as indicated by the arrows (colors). (d)-(e)-(f) show the corresponding net critical current enhancement $\Delta J_C$ associated to presence of the F layer, for the different field sweeps (same color code) shown in (a)-(b)-(d). $\Delta J_C$ was calculated as detailed in the main text from measuremens of the critical current at $T$=3.5 K. The insets show examples of the critical current vs. applied field $J_C(H)$ measured for the three studied samples after demagnetization (blue) and the reference $Mo_{80}Si_{20}$ films (green). Measurement were done at T=3.5K. The critical current was obtained from V(I) characteristics using a voltage criterion $V_C$ = 5 μV ($E_C$ = 25 mV.m$^{-1}$). (h)-(i)-(j) show $\Delta J_C$ calculated with the theoretical model (using the measured $m_z$) discused in the text. The same color code as in (a)-(b)-(c) indicates the different field sweeps.

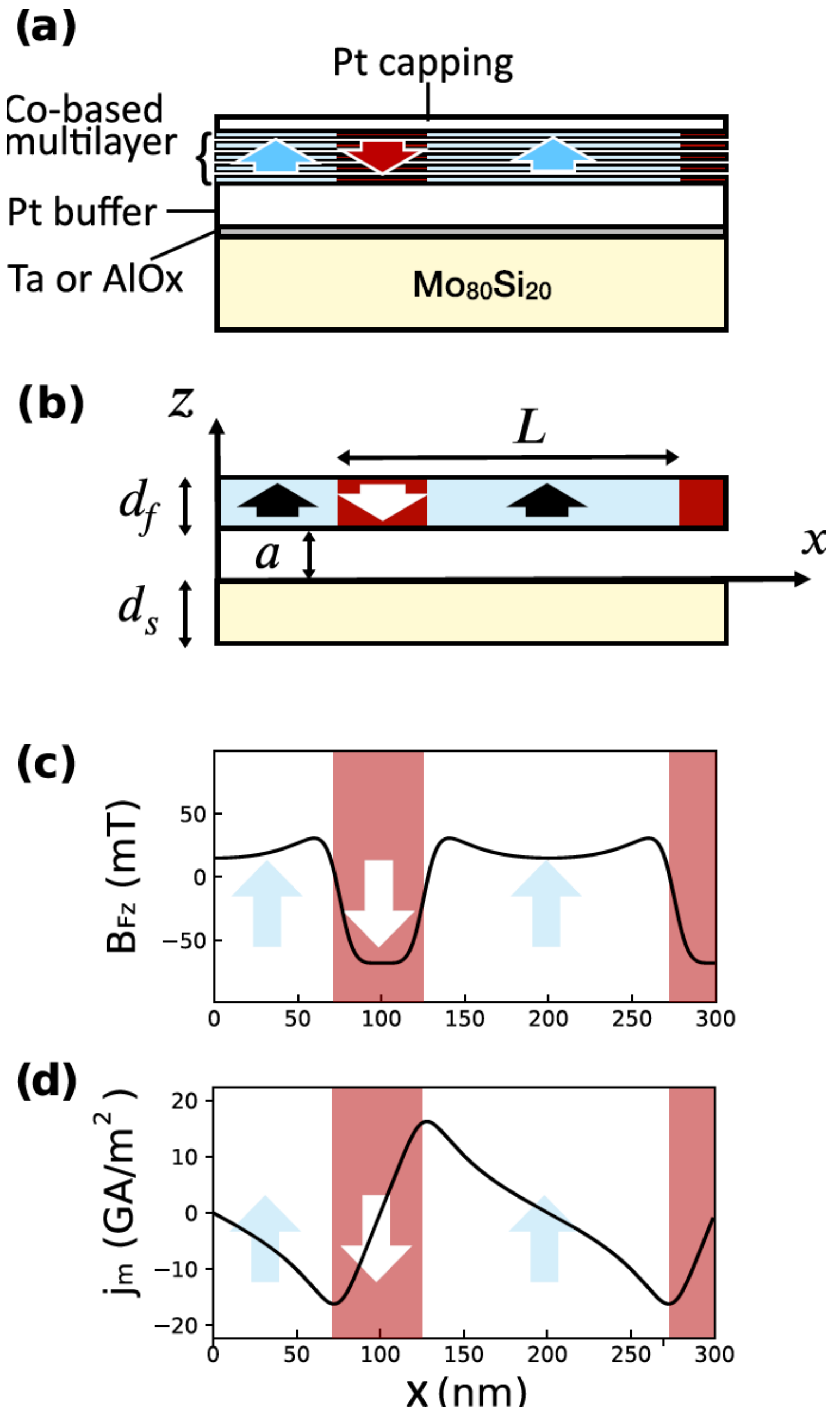

**Figure 4:** (a) Schematic of the model, with a superconductor of thickness $d_s$ and a ferromagnet of total Co thickness $d_{\mathrm{f}}$. The ferromagnet and superconductor are separated by a distance $a$, representing the buffer and insulating layers. Arrows and colors represent the magnetization orientation within each domain. (b) Stray field produced by the domains at the surface of the superconductor. (c) Meissner current induced in the superconductor, calculated for $M_s$ = 1.0 MA.m $^{-1}$ ; $d_f$ = 3nm ; Λ=1200 nm ; $a$=14 nm .